\documentclass[twocolumn]{article}

\usepackage[obeyspaces]{xurl}
\usepackage{hyperref}
\usepackage{newtxtext,newtxmath}
\usepackage[textwidth=515pt, textheight=690pt]{geometry}
\usepackage[T1]{fontenc}
\usepackage{natbib}
\DeclareRobustCommand{\VAN}[3]{#2}
\let\VANthebibliography\thebibliography
\def\thebibliography{\DeclareRobustCommand{\VAN}[3]{##3}\VANthebibliography}

\usepackage{graphicx}	
\usepackage{amsmath}	

\usepackage[
singlelinecheck=false 
]{caption}

\title{\bf{Diffuse $\gamma$-ray emission in Cygnus X: Comments to \\Fornieri \& Zhang 2022} 
}

\author{
Huirong Yan$^{1,2}$\thanks{E-mail: huirong.yan@desy.de}  , 
Parth Pavaskar,$^{1,2}$
\\
\small $^{1}$Deutsches  Elektronen  Synchrotron  (DESY),  Platanenallee  6,  D-15738  Zeuthen,  Germany\\
\small $^{2}$Institut f{\"u}r Physik  und  Astronomie,  Universit{\"a}t Potsdam, D-14476 Potsdam,  Germany\\
}

\date{\today}

\begin{document}

\raggedbottom

\twocolumn[
\begin{@twocolumnfalse}
\maketitle
\begin{abstract}
{\normalsize High energy emissions near particle accelerators provide unique windows to probe the particle acceleration and ensuing escape process determined by local medium properties, particularly the turbulence properties. It has been demonstrated both theoretically and observationally that particle diffusion in local environment can differ from the averaged values inferred from the CR global propagation in the Galaxy determined by local medium, particularly magnetic field and turbulence.
A recent publication by \cite{FZ22} computed particle transport employing the formalism of fast modes scattering calculation from \cite{YL08} and the MHD modes composition results from \cite{MY20} and \cite{Zhang20NA}. The authors claim that the Cygnus X observations from HAWC and Ferm-LAT can be reproduced \citep{Abeysekara:2021yum, Ackermann:2011lfa}. 
\bf{We clarify in this paper that the particle diffusion coefficients they obtained do not correspond to their adopted turbulence and medium properties and are incorrect by an order of magnitude. We also point out that the injection process also plays an indispensable role in determining the high energy particle distribution and the resulting gamma-ray emission.}
\\

{\bf Keywords.} particle transport -- turbulence -- methods: numerical -- diffuse gamma-ray emission -- (magnetohydrodynamics) MHD
}
\end{abstract}
\end{@twocolumnfalse}
\vspace{1.5cm}
]

%



\section{Introduction} \label{sec:intro}

Particle propagation is determined by the properties of interstellar turbulence. The multiphase nature of interstellar medium (ISM) and diversity of driving mechanisms give rise to spatial variation of turbulence properties. Important progress made in the past two decades in cosmic ray propagation is the recognition of inhomogeneous diffusion revealed both from the theoretical understanding of magnetohydrodynamic (MHD) turbulence as well as observations.
The transport of cosmic rays after being released from the acceleration sites determines the diffuse gamma-ray emission. Given the fact that turbulence properties are highly environment dependent, the inhomogeneous cosmic ray diffusion has to be taken into account. Specifically, damping affects the compressible turbulence significantly on small scale \citep{YL02}. Even the isotropy of turbulence can change because of directional damping dependence \citep{YL04, YL08}. On the other hand, injection process determines, in particular, the energy partition among the MHD modes. Understandably, solenoidal driving generates more Alfv\'en modes, whereas compressible driving channels more energy into compressible modes \citep{MY20}. These impacts on the compressible mode energies, particularly fast modes, significantly affect local cosmic ray transport properties. It has been clearly shown from latest observations that the cosmic ray diffusion is neither homogeneous nor isotropic. For instance, the HAWC observation has shown that the diffusion around the Geminga region is two orders of magnitude lower than the average interstellar value \citep{HAWC_Geminga}.

The paper by \cite{FZ22} (hereafter FZ22) applied the properties of observed compressible MHD turbulence to model particle transport and the resulting gamma-ray emission from the Cygnus X region. Although the authors were presented with the problem, and had started their exploration as part of a work assignment at H. Yan’s research group, the scientific disagreement emerged at a later stage, as will be discussed further, on the calculation of the diffusion coefficients. It has occurred to us that the authors treated the damping of fast modes with an ad hoc parameter, instead of calculating it self-consistently from the medium properties.

As a result, it is evident that the calculations presented in the paper have some intrinsic flaws. Unfortunately, the authors have failed to reach an agreement on such discussions. Since the paper employs the results and the expression derived in our previous publications \citep{YL02, YL04, YL08, MY20, Zhang20NA}, we feel the need to inform the community about the inadequacies involved in the analysis in FZ22. ­

\section{Incorrect calculation of the diffusion coefficient in FZ22}

\begin{figure}
 \centering
    \includegraphics[width=8.5cm]{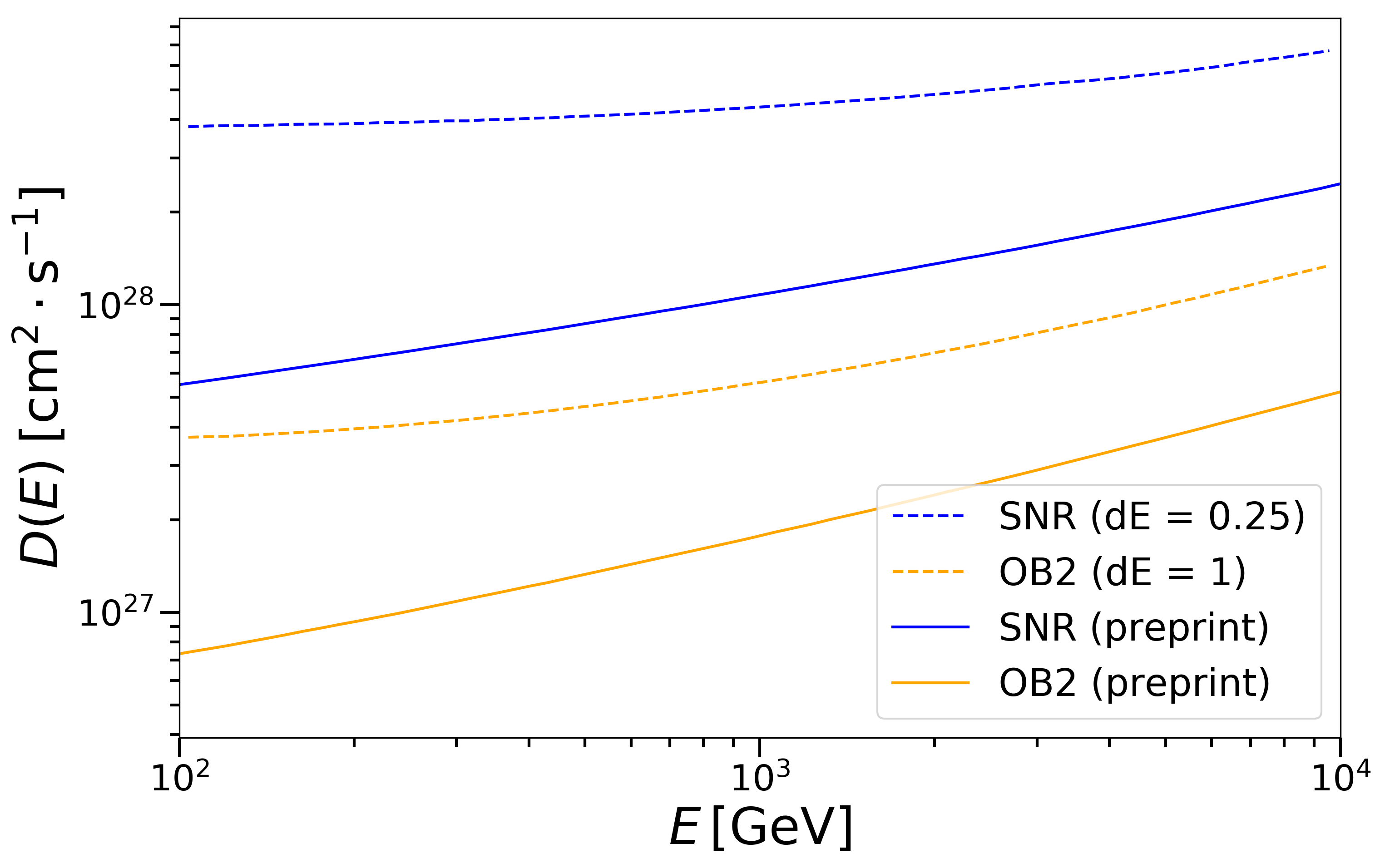}
    \caption{The comparison between the diffusion coefficients adopted in FZ2022 and the ones calculated from scattering in fast modes turbulence with the same input environment parameters as used in FZ2022. dE is the normalized turbulence amplitude, i.e., $\sim M_A^2$, where $M_A$ is the Alfv\'enic Mach number.}
    \label{fig:diff_coef}
\end{figure}

The essence of the paper is to introduce two different diffusion zones based on the degree of turbulence compressibility. As demonstrated by \cite{YL02, YL04,YL08}, cosmic ray scattering is dominated by fast modes. The authors used the formulae from \cite{YL08} without a proper understanding of the physical meaning of the input plasma parameters, particularly the damping of fast modes, a crucial process determining the fast modes angular distribution on small scales. As a result, the particles' diffusion coefficients, which are the basis of the paper, are incorrect. We note that the energy dependence is also not accurate. Adopting the same medium parameters as in FZ22, we recalculated the diffusion coefficients. The comparison is given in Fig.\ref{fig:diff_coef}. The solid lines show the diffusion coefficient obtained in FZ22. The dashed lines represent the diffusion coefficients we calculated using the same environmental parameters as in the paper, which is an order of magnitude larger and the energy dependence is also shallower. 

\section{The role of injection}

\begin{figure*}
    \includegraphics[width=19cm]{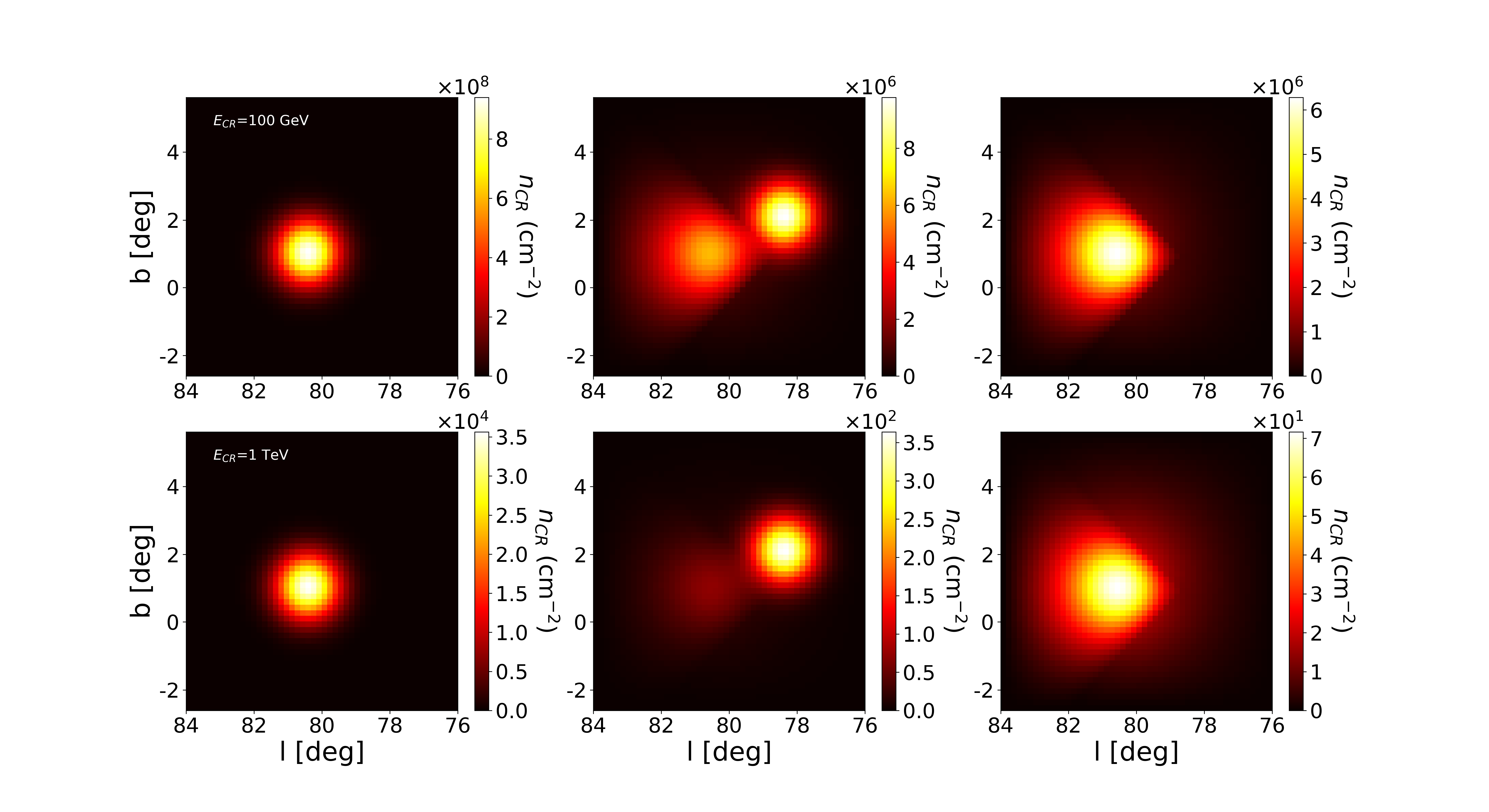}
    \caption{Continuous injection of CR from OB cluster following an initial burst, evolved with with different diffusion coefficients (similar setup to FZ22). The colorbar represents estimated LOS integrated number density of CR protons assuming a standard type-II supernova CR energy budget of $\sim 10^{50}$ erg (10\% of total kinetic energy) and an energy power law with the peak at 1 GeV and a slope of -2.3.}
    \label{fig:diff_injection_cont}
\end{figure*}

\begin{figure*}
    \includegraphics[width=19cm]{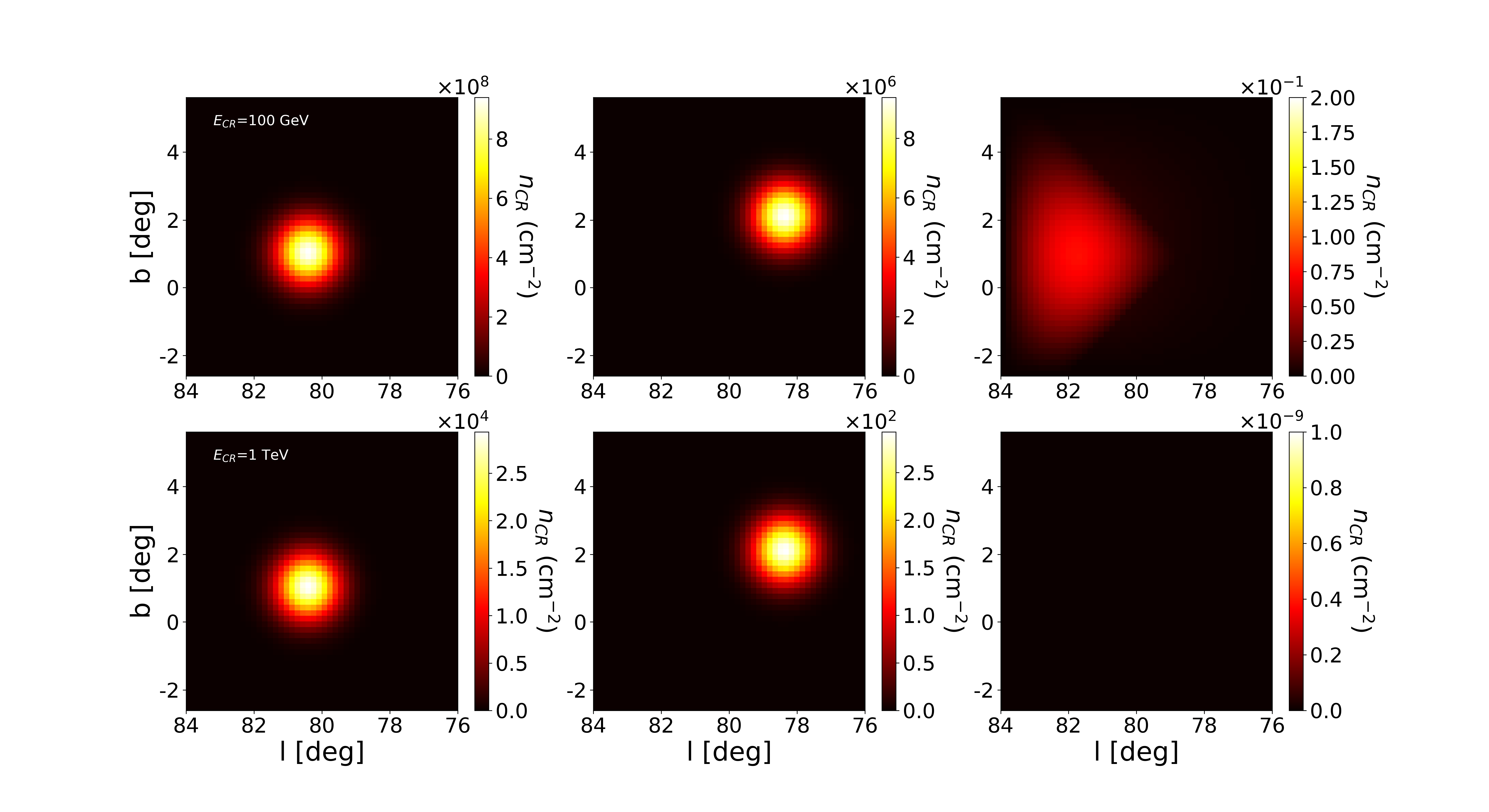}
    \caption{Similar to Fig.\ref{fig:diff_injection_cont} but CR injection is considered as a singe burst at the initial time step without any sustained injected.}
    \label{fig:diff_injection_burst}
\end{figure*}

Since it is not described how the particle distribution in FZ22 is obtained, we ran similar numerical tests with multiple diffusion setups and different particle injection profiles. In the first scenario, we consider a sustained continuous injection following an initial burst at the first time step which is $\approx 1.2$ Myr ago (Fig.\ref{fig:diff_injection_cont}). In the second case, the CR are injected only from the single burst from the OB cluster (Fig.\ref{fig:diff_injection_burst}). 

It is obvious that the first case reproduces the particle distribution from Fig. 2 in FZ22. It can also be seen that the impact of SNR burst is negligible, and that the continuous injection from OB2 is the significant contributor to the resulting particle distributions. In the second case, we see that the particles have diffused completely by the last time step (present-day) despite the secondary injection from the SNR at a much later age. 

\begin{figure*}
 \centering
    \includegraphics[width=19cm]{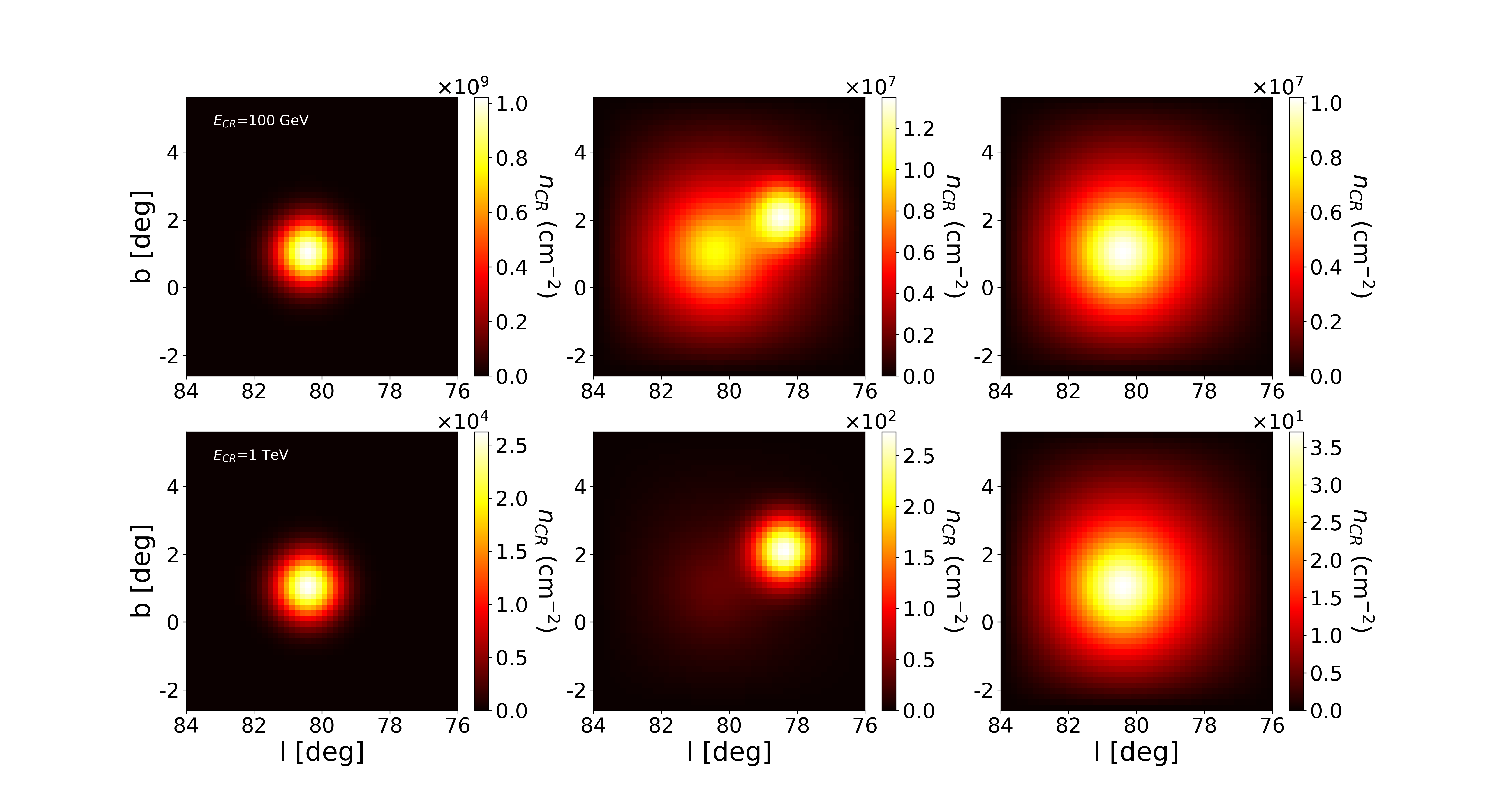}
    \caption{Similar to Fig.\ref{fig:diff_injection_cont}, but with homogeneous diffusion (Continuous injection following initial burst).}
    \label{fig:same_diff_cont}
\end{figure*}

\begin{figure*}
 \centering
    \includegraphics[width=19cm]{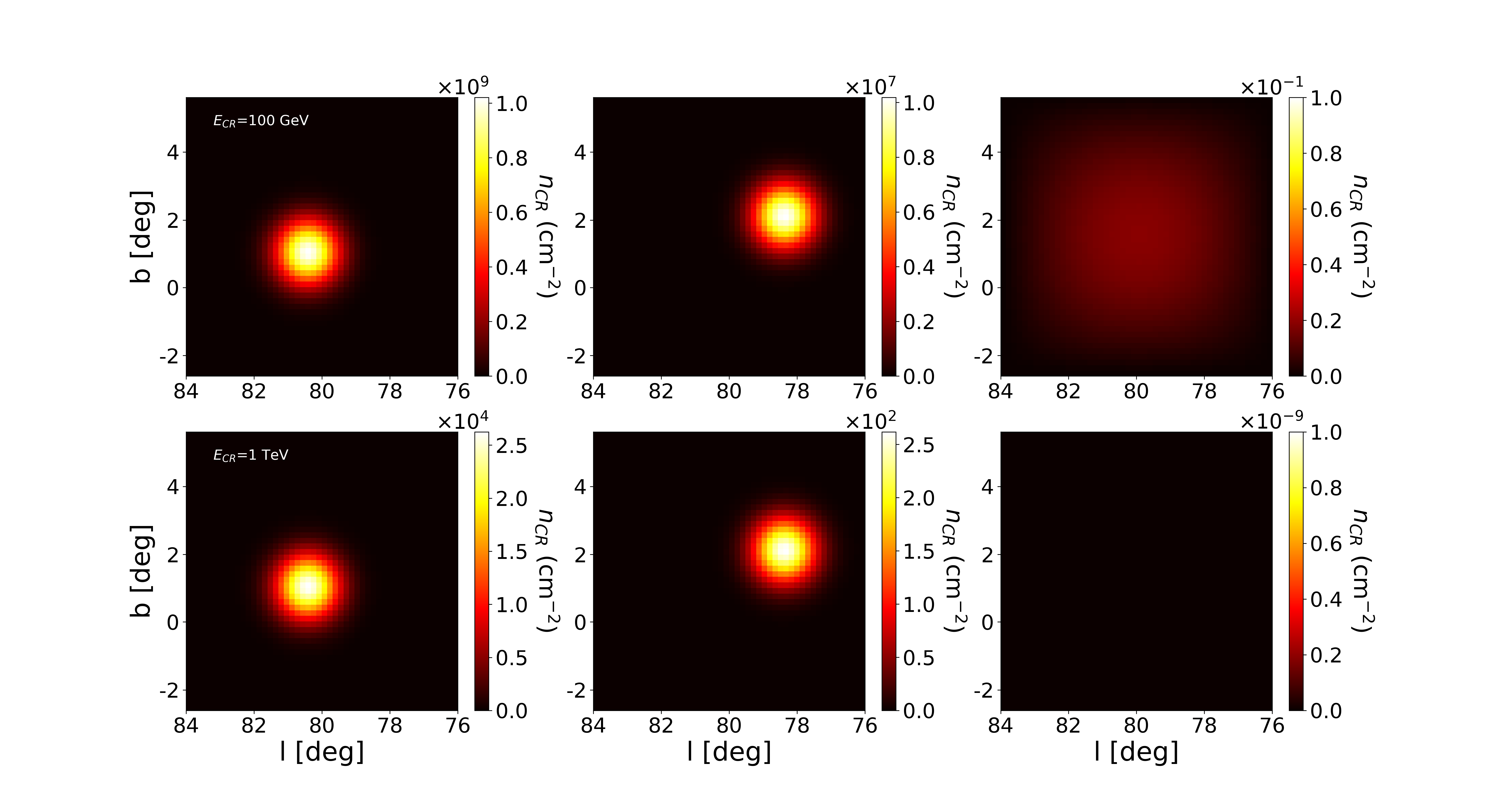}
    \caption{Similar to Fig.\ref{fig:diff_injection_burst}, but with homogeneous diffusion (CR injection from single burst at $t=t_1$).}
    \label{fig:same_diff_burst}
\end{figure*}

\begin{figure*}
    \includegraphics[width=20cm]{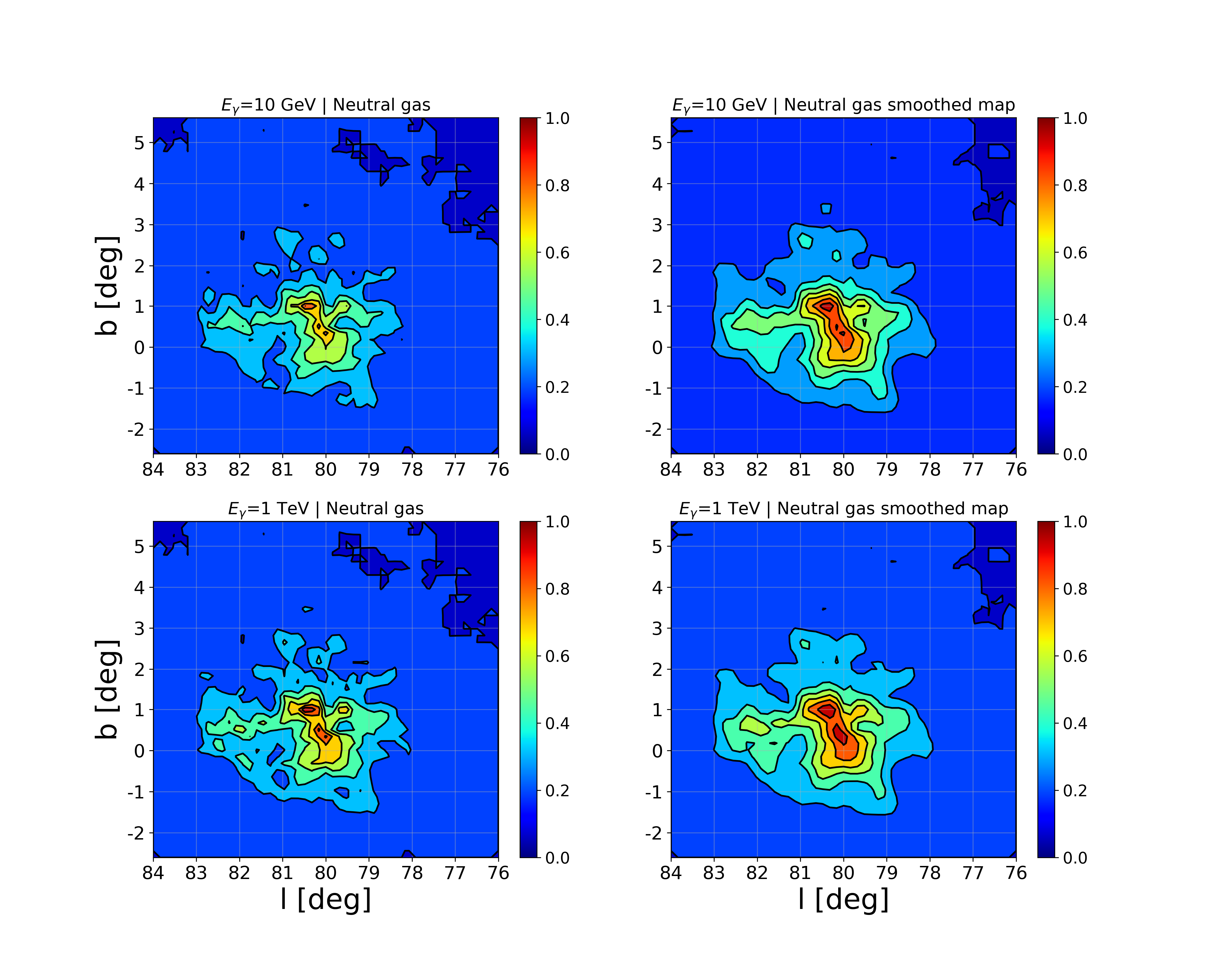}
    \caption{Top: $\gamma$-ray map at E$_\gamma \approx10$ GeV (top) and E$_\gamma \approx1$ TeV (bottom) generated with the original CO map (left) and the smoothed CO map (right). The CR distribution obtained from the homogeneous diffusion model is used for the convolution with the CO map.}
    \label{fig:gamma_map}
\end{figure*}

On the other hand, we also computed the particle distribution with the same injection setup, but with homogeneous diffusion. Fig.\ref{fig:same_diff_cont} and Fig. \ref{fig:same_diff_burst} show that the present-time particle densities are also different than the earlier case of inhomogenous diffusion. As in the two zone diffusion case, only continuous injection leaves non-negligible particle concentration in the present day (Fig.\ref{fig:same_diff_cont}). In the case of continuous injection corresponding to Fig.\ref{fig:same_diff_cont}, we computed the gamma-ray map following the procedure described in FZ22. Fig.\ref{fig:gamma_map} demonstrates that the expected gamma-ray emission in the case of homogeneous diffusion differs from both Fermi and HAWC observations \citep{Abeysekara:2021yum, Ackermann:2011lfa}. Our results suggest that both the spatial inhomogeneity of the diffusion and the injection mechanism play critical roles in the particle distributions.

\section{Summary}

Local magnetic field and turbulence properties are vital to study the high energy emission, particularly diffuse ones. While the efforts of using first-hand knowledge of turbulence to model high energy phenomena are generally welcome in the community, the paper by \cite{FZ22} disqualifies itself from making any robust conclusions owing to the incorrect calculation of diffusion coefficients based on local turbulence parameters. 
Their adopted diffusion coefficient is 1-2 orders magnitude smaller than the interstellar value.
From observational fittings, we also find that {\bf such slow diffusion is unnecessary} (different from the case of Geminga), although a two-zone diffusion model is preferable. Moreover, the particle injection mechanisms also play a decisive role and have to be carefully studied in order to understand the observations.


\section*{Data Availability}

The data involved in this work will be shared upon reasonable request to the corresponding author.

\section*{Acknowledgments}  HY \& PP acknowledge the help from SQ Zhao, who performed an independent check on the diffusion calculation. PP would like to thank S. Malik for the helpful discussions.




\bibliographystyle{aasjournal}
\bibliography{CRPaper}

\end{document}